\documentclass[reprint,aps,amsmath,amssymb,twoside,prd,showkeys,superscriptaddress]{revtex4-1}
\pdfoutput=1
\usepackage{verbatim}
\usepackage{comment}
\usepackage{graphicx}
\usepackage[T1]{fontenc}
\usepackage{epsfig}
\usepackage{bm}
\usepackage{amssymb}
\usepackage{float}
\usepackage{amsmath}
\usepackage{subfigure}
\usepackage{dcolumn}
\usepackage{cancel}
\usepackage[colorlinks]{hyperref}
\usepackage[usenames,dvipsnames]{color}
\hypersetup{
     breaklinks=true,
    pdfstartview={FitH},    
    colorlinks=true,       
    linkcolor=blue,          
    citecolor=red,        
    filecolor=magenta,      
    urlcolor=blue,           
    anchorcolor=green,      
    linktocpage=true
}

\def\doi{http://doi.org}

\newcommand{\HCd}{\mathcal{H}}

\def\HCdt0{\tilde{\HCd}_{0}}

\newcommand{\onehalf}{{\textstyle\frac{1}{2}}}

\newcommand{\affcam}{DAMTP, Centre for Mathematical Sciences, University of Cambridge, Wilberforce Road, Cambridge CB3 0WA, UK}
\newcommand{\affkicc}{Institute of Astronomy, University of Cambridge, Madingley Road, Cambridge, CB3 0HA, UK}

\begin{document}
\title{Cosmology of fermionic dark energy coupled to curvature}
\author{David Benisty}
\email{db888@cam.ac.uk}
\affiliation{\affcam}\affiliation{\affkicc}
\begin{abstract}
A formulation of cosmology driven by fermions $\psi $ is studied. Assumption
that the expectation value of the fermion bilinear is non-zero simplifies
the homogeneous solution of the Dirac equations and connects the spinor
field with the scale parameter of the universe. With coupling between the
Einstein term and spinor field
$1 - \frac{\xi }{6} (\bar{\psi }\psi )^{-l}$, the possibility for a late
time interaction emerges. In that way, the early universe agrees with
$\Lambda $CDM model, but for the late universe the new integrating term
dominates. 
\end{abstract}
\maketitle
\section{Introduction} 
The standard model of cosmology, the $\Lambda$CDM model, requires a dark energy (DE) component responsible for the observed late-time acceleration of the expansion rate. The tension between the values of the Hubble constant $H_0$ obtained from the late universe measurements \cite{Riess:2019cxk,Riess:2020fzl} and those from the Cosmic Microwave Background (CMB) by Planck Collaboration \cite{Aghanim:2018eyx} is larger than $4\sigma$. This tension is one of the biggest challenges in modern cosmology that indicates a possible new physics \cite{DiValentino:2020vhf,DiValentino:2020zio,DiValentino:2020vvd,Efstathiou:2020wxn,Borhanian:2020vyr,Hryczuk:2020jhi,Klypin:2020tud,Ivanov:2020mfr,Chudaykin:2020acu,Lyu:2020lps,Alestas:2020mvb,Motloch:2019gux,Frusciante:2019puu,Kumar:2021eev,DiValentino:2021izs,Alestas:2020zol,Benisty:2021wxi}.

The search for constituents responsible for accelerated periods in the evolution of the universe is a big topic in cosmology. Several candidates has been tested for
describing both the inflationary period and the present
accelerated era: scalar fields, equations of state
and cosmological constants. Fermionic fields has also been tested as gravitational sources of an expanding universe \cite{Samojeden:2010rs} with cosmological applications ~\cite{Chimento:2007fx,Myrzakulov:2010du,Ribas:2016ulz}. Fermions as dark matter are widely discussed such as neutrino dark matter, or supersymmetric extensions of the Standard Model \cite{Roszkowski:1993by,Roulet:1993xs,Drees:1994ut,Jungman:1995df,Martin:1997ns,Bertone:2004pz,Feng:2010gw,Ellis:2010kf}. In some of these models the fermionic field plays the role of the inflaton in the early period of the universe and of dark energy for the late universe \cite{Benisty:2019jqz}.

As we will see, under the simplest assumption of an homogeneous expansion the Noether symmetry yields the relation $\bar{\psi}\psi \sim 1/a^3$ where $\psi$ is the fermionic field, and $a$ is the scale factor of the universe. This relation impose a different scenario than the scalar fields in cosmology and gives another candidates for dark matter and dark energy \cite{Zee:1979hy,Cooper:1981byv,Rossi:2019lgt,Ballardini:2020iws,Ballesteros:2020sik}. In the inflationary scenario the scalar fields need to be slow roll, which mean that for $60$ e-folds the scalar field value should be around the same value $\phi \approx Const$ \cite{Starobinsky:1979ty,Kazanas:1980tx,Starobinsky:1980te,Guth:1980zm,Linde:1981mu,Albrecht:1982wi,Barrow:1983rx,Blau:1986cw,Barrow:2016qkh,Barrow:2016wiy,Olive:1989nu,Linde:1993cn,Liddle:1994dx,Lidsey:1995np,CervantesCota:1995tz,Berera:1995ie,ArmendarizPicon:1999rj,Kanti:1999ie,Garriga:1999vw,Gordon:2000hv,Bassett:2005xm,Chen:2009zp,Germani:2010gm,Kobayashi:2010cm,Feng:2010ya,Burrage:2010cu,Kobayashi:2011nu,Ohashi:2012wf,Hossain:2014xha,Hossain:2014zma,Cai:2014uka,Geng:2015fla,Kamali:2016frd,Geng:2017mic,Benisty:2017lmt,Dalianis:2018frf,Dalianis:2019asr,Benisty:2020vvm,Benisty:2019pxb,Benisty:2019bmi,Benisty:2019tno,Staicova:2019ksr,Staicova:2018ggf,Staicova:2018bdy,Guendelman:1999rj,Guendelman:2014bva,Qiu:2020qsq}. However, the corresponding fermionic fields evolves $e^{-180}$ times. Therefore, any similarity between scalar and fermions is only accidental.

The corresponding theory that involve fermions instead of scalars are the "Fermions Tensor Theories" (FTT). Here we show that a simple coupling to curvature gives an exaltation to $H_0$ tension. This new coupling is dominant for the late universe and gives closer value to local measurement of the Hubble constant from \cite{Riess:2019cxk}. For the early universe we get the $\Lambda$CDM model.

The plan of the work is as follows: Section \ref{sec:fercur} formulates the tetrad formalism for a homogeneous expansion. Section \ref{sec:fertenthe} suggests the FTT with the solved equations of motions. Section \ref{sec:decCoup} gives the extended theory we test. Section \ref{sec:obs} test the extended model with the latest observations. Section \ref{sec:res} summarizes the results that finalizes in \ref{sec:dis}.

\section{Fermions in Curved Spacetime} 
\label{sec:fercur}
In this section we construct the scenario of fermionic cosmology in FTT. We first briefly review the basics of fermions in curved spacetime, and then we present the Lagrangian of the model, extracting the equations of motion. Fermions in a cosmological background have been widely discussed in \cite{Saha:2019ztr}. The tetrad formalism was used to combine the gauge group of general relativity with a spinor matter field. The tetrad $e^{a}_{\mu}$ and the metric $g_{\mu\nu} $ tensors are related through
\begin{equation}
g_{\mu\nu} = e^{a}_{\mu} e^{b}_{\nu} \eta_{a b}, \quad a,b = 0,1,2,3,
\label{metrictetrad}
\end{equation} 
with Latin indices refer to the local inertial frame  with the Minkowski metric $\eta_{ab}$, while Greek indices denote the local coordinate basis of the manifold.

$\gamma^a$ are the Dirac matrices in the standard representation (flat spacetime). The Dirac matrices in curved space $\Gamma^\mu = e_a^{\phantom{a} \mu} \gamma^a$ are obtained using the tetrads  $e_a^{\phantom{a} \mu}$, labeled with a Latin index. The generalized Dirac matrices obey the Clifford algebra $\{\Gamma^{\mu},\Gamma^{\nu}\} = 2 g^{\mu\nu}$. The definition for the covariant derivative for spinors reads:
\begin{equation}
\label{1a}
\psi_{;\mu}= \partial_\mu\psi-\Omega_\mu\psi, \quad
\overline\psi_{;\mu}=\partial_\mu\overline\psi+\overline\psi\Omega_\mu,
\end{equation}
\begin{equation}
\Omega_\mu=\frac{1}{4}g_{\beta \nu} \left[\{ \genfrac{}{}{0pt}{}{\nu}{\alpha \mu} \} - e^{\nu}_{j} 
\partial_{\mu} e^{j}_{\alpha} \right] \gamma^{\beta} \gamma^{\alpha},
\end{equation}
where $\Omega_\mu$ is the spin connection, and $\{ \genfrac{}{}{0pt}{}{\nu}{\alpha \mu} \}$ is the Christoffel symbols:
\begin{equation}
\{ \genfrac{}{}{0pt}{}{\rho}{\mu \nu}\} = \onehalf g^{\rho\lambda} (g_{\lambda\mu,\nu}+g_{\lambda\nu,\mu}-g_{\mu\nu,\lambda}),
\end{equation}
We consider the Friedman Lema\^tre Robertson Walker (FLRW) homogeneous and isotropic metric:
\begin{equation}\label{eq:metric}
ds^2 =-n^2 dt^2 +a^2 \left(dx^2+dy^2+dz^2\right),
\end{equation}
with the scale factor $a$ and the Lapse function $n$. Through (\ref{metrictetrad}) the tetrad
components are found to be
\begin{equation}
e^{\mu}_{0} = \frac{1}{n}\delta^{\mu}_{0}, \quad  e^{\mu}_{i} = \frac{1}{a}\delta^{\mu}_{i}.
\end{equation}
Moreover, the covariant version of the Dirac matrices are
\begin{equation}
\Gamma^{0} = \frac{1}{n}\gamma^{0}, \quad \Gamma^{i} =\frac{1}{a}\gamma^{i},
\end{equation}
while the spin connection becomes
\begin{equation}
\Omega_{0} = \gamma^{0}, \quad \Omega_i = \frac{1}{2} \frac{\dot{a}}{n} \gamma^{i}\gamma^0.
\end{equation}
$\dot{a}$ = da/dt, is the time derivative. This formulation is the basic mathematics for the Fermion tensor theories (FTT).
\section{Fermion Tensor Theories}
\label{sec:fertenthe}
 {In this section we apply the above fermionic formulation at a cosmological framework, focusing on late-time universe. The framework of FTT has a similar form to Scalar Tensor Theories and discussed in details in \cite{Saha:2019ztr,deSouza:2008az}.} The FTT read:
\begin{equation}
\begin{split}
\mathcal{L} =  \frac{f(\phi)}{2\kappa^2}  \mathcal{R} + \frac{i}{2}\left[ \bar{\psi} \Gamma^\mu
\psi_{;\mu} - \bar{\psi}_{;\mu} \Gamma^\mu \psi \right] -  V(\phi) + \mathcal{L}_m, \label{eq:Lag} 
\end{split}
\end{equation}
where $ \mathcal{R}$ is the Ricci scalar, $\psi$ and $\overline \psi=\psi^\dag \gamma^0$ are the spinor field and its adjoint, respectively.  We set $\kappa^2 = 1/8 \pi G = 1$. The scalar $\phi \equiv |\bar\psi \psi| $ multiples the fermionic field and its conjugate field. $f$ is the function that couples the Einstein term and $ V $ the self-interaction potential density of the fermionic field. The kinetic term of the spinors is the same as the kinetic term  from Dirac equation. However, FTT suggest a generic coupling function $f$.

For $f = 1$ and $V = m_{\psi} \phi$ the FTT reduce to the Dirac equation in curved spacetime. For the metric (\ref{eq:metric}) the action (\ref{eq:Lag}) reduces to the form:
\begin{equation}
\begin{split}
\mathcal{L} = \frac{1}{2} a^2 \bar{\psi} \psi \dot{a}-\Lambda  a^3 n-a^3 n V\left(\bar{\psi} \psi\right)\\
+f\left(\bar{\psi} \psi\right) \frac{a  \left(6 n \left(a \ddot{a}+\dot{a}^2\right)-6 a \dot{a} \dot{n}\right)}{2 n^2} + n a^3 \mathcal{L}_m
\end{split}
\label{eq:LagMini}
\end{equation}
$\dot{\psi} = \frac{d}{dt}\psi$ is the time derivative. The solution is obtained via the complete set of variations: the scale factor $a$ and the spinor field $\psi$. The variation with respect to field $\psi$ (extended Dirac equation) yields: 
\begin{equation}
\dot{\psi }+ \frac{3}{2} H \psi + i \gamma^{0}\psi V'(\phi) - 6 i (\dot{H}+ 2 H^2 ) \gamma^0 \psi f'(\phi) = 0,
\label{gendirac1}
\end{equation}
and similarly for $\bar{\psi}$:
\begin{equation}
\dot{\bar{\psi} }+ \frac{3}{2} H \bar{\psi} - i \bar{\psi} \gamma^{0} V'(\phi) + 6 i \bar{\psi} (\dot{H}+ 2 H^2 ) \gamma^0 f'(\phi) = 0.
\label{gendirac2}
\end{equation}
$H(t)$ is the Hubble function: $H(t) = \frac{\dot{a}}{a}$. By multiplying the first equation by $\bar{\psi}$ from the left hand side and the second equation by $\psi$ from the right hand, the equations get the same form. The sum of those two equations gives:
\begin{equation}
\frac{d}{dt}\left(\bar{\psi}\psi\right) + 3 H \, \bar{\psi}\psi= 0,
\end{equation}
with the solution: \begin{equation}
\phi := \bar{\psi}\psi = n_\psi/a^3.
\label{eq:psipsibar}
\end{equation}
$n_\psi$ is a dimensionless integration constant that emerges from Eq (\ref{eq:psipsibar}). The quantity $m_\psi n_\psi$ is thus the total gas energy density. The variation w.r.t the Lapse function yields the Friedmann's equation:
\begin{equation}\label{eq:Friedmann}
3 H^2  = \frac{V(\phi) + \rho_m}{f(\phi)-3\phi f'(\phi)},
\end{equation}
with the  {gauge} $n(t) = 1$. Together with the set of Eq. (\ref{eq:psipsibar}) one can find the whole evolution of the model.
\begin{figure}[t!]
 	\centering
\includegraphics[width=0.45\textwidth]{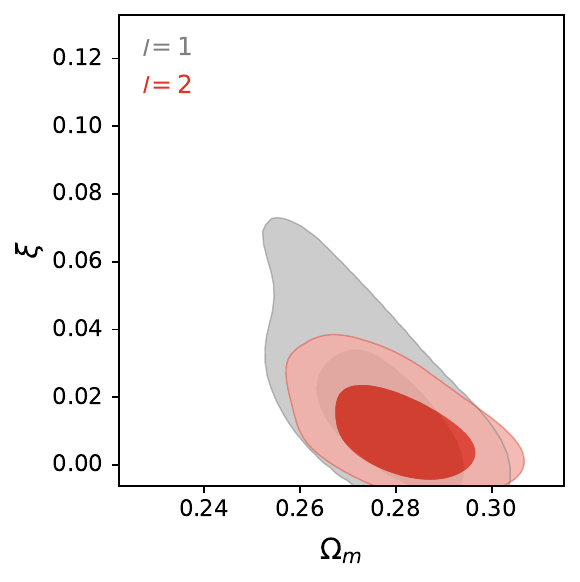}
\caption{\it{The posterior distribution for different measurements with the $\Lambda$CDM model with $1 \sigma$ and $2\sigma$. The gray contour describes the model with $l = 1$, The red contour describes the model with $l = 2$.}}
 	\label{fig:1}
\end{figure}

\section{Decaying curvature coupling} 
\label{sec:decCoup}
The $\Lambda$CDM model emerges from the simplest expression for the potential. A simple example is a linear expansion for both functions:
\begin{equation}
f(\phi) = 1, \quad V(\phi) =  m_\psi  \phi + \Lambda,
\end{equation}
 {with the dimensions: $[ m_\psi ] = L^{-1}.$} The corresponding Hubble function reads:
\begin{equation}
H^2/H_0^2 = \Omega_m/a^3 + \Omega_\Lambda,
\end{equation}
with the dark matter component 
\begin{equation}
 \Omega_m = \frac{m_{\psi} n_{\psi}}{3 H_0^2}, \quad    \Omega_\Lambda =  \frac{\Lambda}{3 H_0^2}
\end{equation}
without other fields. $H_0$ is the predicted value of  {at present time}. In this model the  {matter} component comes from the fermions field. \cite{deSouza:2008az} uses the Noether symmetry in order to find  the coupling function $f$. The paper suggests $f \sim \phi$ or $f \sim \phi^{1/3}$. The case of $f \sim \phi^{1/3}$ gives a divergent solution. Inspired by \cite{deSouza:2008az} solutions, we choose the following combinations, with a power-law extension:
\begin{equation}
f(\phi) = 1 - \frac{\xi}{6}\phi^{-l}, \quad V(\phi) =  m_\psi  \phi,
\end{equation}
The corresponding Hubble function yields:
\begin{equation}
\left(\frac{H(z)}{H_0}\right)^2 = \frac{\Omega_\Lambda  + \Omega_m (1+z)^3 + \Omega_r (1+z)^4 }{1 + \tilde{\xi}/ (1+z)^{3l}},
\end{equation}
with: $\tilde{\xi} = \left( 3l - 1\right)\xi/6 n_{\psi}^{l}$. The parameter $\tilde{\xi}$ is combined from the parameters $n_{\psi}$, $l$ and $\xi$. the parameter $\tilde{\xi}$ is a dimensionless parameter. Because of the ambiguity with the other parameters we use the $\tilde{\xi}$ parameter instead. The new  {parameterization} changes the behavior of the late universe, since for large redshifts the term $\tilde{\xi}/ (1+z)^{3l}$ decays. Hence the term "Decaying coupled Fermions to curvature". For the future expansion the model predicts different behavior.

\section{Observational Constraints} 
\label{sec:obs}
In the following we describe the observational data sets along with the relevant statistics in constraining 
the model.

\subsubsection{Direct measurements of the Hubble expansion}
Cosmic Chronometers (CC): The data set exploits the evolution of differential ages of passive galaxies at different redshifts to directly constrain the Hubble parameter \cite{Jimenez:2001gg}. We use uncorrelated 30 CC measurements of $H(z)$ discussed in \cite{Moresco:2012by,Moresco:2012jh,Moresco:2015cya,Moresco:2016mzx}. Here, the corresponding $\chi^2_{H}$ function reads:
\begin{equation}
\chi^{2}_{H} = \sum_{i=1}^{30}\left(\frac{H_{i} - H_{pred}(z_i)}{\Delta H_i}\right)^2,
\end{equation}
 {where $H_{i}$ is the observed Hubble rates at redshift $z_{i}$ ($i=1,...,N$) and $H_{pred}$ is the predicted one from the model.}

\begin{figure*}[t!]
 	\centering
\includegraphics[width=0.9\textwidth]{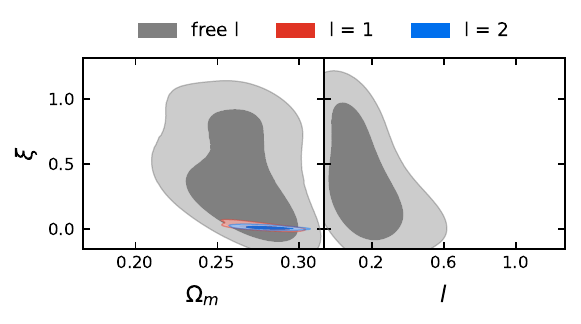}
\caption{\it{The posterior distribution for different measurements with the $\Lambda$CDM model with $1 \sigma$ and $2\sigma$.}}
\tabcolsep 5.5pt
\vspace{1mm}
\centering
\begin{tabular}{|c|c|c|c|c|} \hline \hline
Parameter & $l = 1$  & $l = 2$ & $l$ free $ = 0.189 \pm 0.385$& $\Lambda$CDM
\vspace{0.05cm}\\ \hline \hline
$\Omega_m$ &  {$ 0.276 \pm 0.013$} & {$0.281 \pm 0.095$} &  {$0.269 \pm 0.018$} &  {$0.286 \pm 0.081$} \\
$ {\tilde{\xi}}$ &  {$\left(2.244\pm 2.087\right) \cdot 10^{-2}$} &  {$\left( 9.51 \pm 8.77\right) \cdot 10^{-3}$} &  {$0.317 \pm 0.578$} & $0$\\
\hline\hline
\end{tabular}
 	\label{fig:2}
\end{figure*}

\subsubsection{Standard Candles}
As Standard Candles (SC) we use measurements of the Pantheon Type Ia supernova \cite{Scolnic:2017caz}. The model parameters of the models are to be fitted with, comparing the observed $\mu _{i}^{obs}$ value to the theoretical $\mu _{i}^{th}$ value of the distance moduli which are the
logarithms:
\begin{equation}
 \mu=m-M=5\log _{10}(D_{L})+\mu _{0},   
\end{equation}
 {where $m$ and $M$ are the apparent and absolute magnitudes and $\mu_{0}=5\log \left( H_{0}^{-1}/Mpc\right) +25$ is the nuisance parameter that
has been marginalized. The luminosity distance is defined by, }
\begin{eqnarray}
D_L(z) &=&\frac{c}{H_{0}}(1+z)\int_{0}^{z}\frac{dz^{\ast }}{%
E(z^{\ast })} ,
\end{eqnarray}%
Here $\Omega_k=0$ corresponds to spatially flat spacetime. Following the approach used in (\cite{DiPietro:2002cz,Nesseris:2004wj,Perivolaropoulos:2004yr,Lazkoz:2005sp}), we assumed no prior constraint on $M_B$, which is just some constant, and we integrated the probabilities over $M_B$. The integrated $\chi^2$ yields:
\begin{equation}
\tilde{\chi}^2_{SN} = D-\frac{E^2}{F} + \ln\frac{F}{2\pi},
\end{equation}
where:
\begin{subequations}
\begin{equation}
D = \sum_i \left(\frac{\mu_{}^{i} - 5 \log_{10}\left[d_L(z_i)\right]}{\sigma_i}\right)^2, 
\end{equation}
\begin{equation}
E = \sum_i \frac{\mu_{}^{i} - 5 \log_{10}\left[d_L(z_i)\right]}{\sigma_i^2},
\end{equation}
\begin{equation}
F = \sum_i \frac{1}{\sigma_i^2}.
\end{equation}
\end{subequations}
Here $\mu_{}^{i}$ is the observed luminosity, $\sigma_i$ is its error, and the $d_L(z)$ is the luminosity distance. The values of $M$ and $H_0$ don't change the marginalized $\tilde{\chi}^2_{SN}$.  {In order to use the covariance matrix provided for the Pantheon dataset, one needs to transform $D,E,F$ as follows:}
\begin{subequations}
\begin{equation}
D = \sum_i \left( \Delta\mu \, C^{-1}_{cov} \, \Delta\mu^T \right)^2,
\end{equation}
\begin{equation}
E = \sum_i \left( \Delta\mu \, C^{-1}_{cov} \, E \right),
\end{equation}
\begin{equation}
F = \sum_i  C^{-1}_{cov}  ,
\end{equation}
\end{subequations}
where $\Delta\mu =\mu_{}^{i} - 5 \log_{10}\left[d_L(z_i)\right)$, $E$ is the unit matrix, and $C^{-1}_{cov}$ is the inverse covariance matrix of the dataset.


\subsubsection{Joint analysis and model selection}
In order to perform a joint statistical analysis of $6$ cosmological probes we need to use the total likelihood function, consequently the  $\chi^2_{\text{tot}}$ expression is given by:
\begin{equation}
\chi_{\text{tot}}^2 = \chi_{H}^2 + \chi_{SC}^2.
\end{equation}
Regarding the problem of likelihood maximization, we use an affine-invariant Markov Chain Monte Carlo sampler \cite{ForemanMackey:2012ig}, as it is implemented within the open-source packaged \textit{Polychord} \cite{Handley:2015fda} with the \textit{GetDist} packages \cite{Lewis:2019xzd} to present the results. The prior we choose is with a uniform distribution, where $\Omega_{m} \in [0.;1.]$, $H_0\in [50;100] \, \text{Km/sec/Mpc}$ and $\tilde{\xi}\in [0;10]$, $l\in [0;3]$ for the additional parameters.

\section{Results}
\label{sec:res}
 {The table under~\ref{fig:2} summarizes the best fit results for different models. Fig~\ref{fig:1},\ref{fig:2} shows the posterior distribution for the $\xi$ vs. $\Omega_m$ and $l$. For the cases the $l$ is assumed to be a constant, i.e. $1$ or $2$, the $\Omega_m$ part slightly changes from $0.286 \pm 0.081$ for the $\Lambda$CDM model: For $l = 1$ the matter part gives $ 0.276 \pm 0.013$ and for $l = 2$ the matter part yields $0.281 \pm 0.095$. For the dynamical $l$ the matter part still gives $0.269 \pm 0.018$. From this comparison we can see the model requires a bit smaller values for the matter part, and does not sensitive to the value of $l$ so much. The conclusion is similar: the model requires a bit smaller values for the dark energy part, and does not sensitive to the value of $l$ by much.}

 {The additional parameters $\tilde{\xi}$ gets different constraints from different models: For $l = 1$ the parameter is constrained to be $\left(2.244\pm 2.087\right) \cdot 10^{-2}$, and for $l = 2$ the parameter is $\left( 9.51 \pm 8.77\right) \cdot 10^{-3}$. For the dynamical $l$ case, the additional parameter is $\tilde{\xi} = 0.317 \pm 0.57$, with $l = 0.189 \pm 0.385$. For these cases it possible to see that $\Lambda$CDM is part from the posterior distribution. For the general case, if we approach to the early universe which referrers to larger $z$, the modified model is changes to $\Lambda$CDM model, since the new term $\left(1+z \right)^{-3l}$ becomes negligible.}

\section{Discussion}
\label{sec:dis}
In this work, we have studied in detail the phenomenology of a simple generalized fermionic field model for dark energy and dark matter.  Assumption that the expectation value of the fermion bilinear is non-zero (which is the case in the presence of dark matter) simplifies the homogeneous solution of the Dirac equations and connects the spinor field with the scale parameter of the universe. With coupling between the Einstein term and spinor field, the possibility for a late time interaction emerges. In that way, the early universe agrees with $\Lambda$CDM model, but for the late universe the new interacting term dominates.

In this model we treat the fermions as the source to the dark matter part. Some plausible option that fits to this model the neutrino dark matter in supersymmetric extensions of the Standard Model. In addition to the matter part, we coupled the fermionic field to the Einstein part using $f(\phi)$ function. Extending the Noether Symmetry that imposes some $f(\phi)$ we set a general form of the function. The form for high redshifts decays, but for small redshifts the model changes from the $\Lambda$CDM.

We test a combine data set of the Cosmic Chronometers and Standard Candles of the Type Ia Supernova. The decaying coupling reveals the capabilities of the scenario and makes it a good candidate for the description of nature. In the future it would be interesting to extended the current work for the perturbation level and to test whenever this kind of coupling may explain the $\sigma_8$ tension while on the homogenous level can keep $\Lambda$CDM for the early Universe. This work should be done with the full analysis of the perturbation solution of the fermionic field.

\acknowledgments
I would like to thank Alexander Kaganovich, Emil Nissimov \& Sventlana Pacheva for helpful comments and advice. This article is supported by COST Action CA15117 "Cosmology and Astrophysics Network for Theoretical Advances and Training Action" (CANTATA) of the COST (European Cooperation in Science and Technology) and the Actions CA16104 and CA18108. Finally I would like to thank to the Grants Committee of the Rothschild and the Blavatnik Cambridge Fellowships for generous supports.

\bibliographystyle{apsrev4-1}
\bibliography{ref}

\end{document}